\documentstyle[12pt,epsfig]{article}
\makeatletter
\def\@cite#1#2{{[{#1}]\if@tempswa\typeout
{IJCGA warning: optional citation argument
ignored: `#2'} \fi}}
\newcount\@tempcntc
\def\@citex[#1]#2{\if@filesw\immediate\write\@auxout{\string\citation{#2}}\fi
  \@tempcnta\z@\@tempcntb\m@ne\def\@citea{}\@cite{\@for\@citeb:=#2\do
    {\@ifundefined
       {b@\@citeb}{\@citeo\@tempcntb\m@ne\@citea\def\@citea{,}{\bf ?}\@warning
       {Citation `\@citeb' on page \thepage \space undefined}}%
    {\setbox\z@\hbox{\global\@tempcntc0\csname b@\@citeb\endcsname\relax}%
     \ifnum\@tempcntc=\z@ \@citeo\@tempcntb\m@ne
       \@citea\def\@citea{,}\hbox{\csname b@\@citeb\endcsname}%
     \else
      \advance\@tempcntb\@ne
      \ifnum\@tempcntb=\@tempcntc
      \else\advance\@tempcntb\m@ne\@citeo
      \@tempcnta\@tempcntc\@tempcntb\@tempcntc\fi\fi}}\@citeo}{#1}}
\def\@citeo{\ifnum\@tempcnta>\@tempcntb\else\@citea\def\@citea{,}%
  \ifnum\@tempcnta=\@tempcntb\the\@tempcnta\else
   {\advance\@tempcnta\@ne\ifnum\@tempcnta=\@tempcntb \else \def\@citea{--}\fi
    \advance\@tempcnta\m@ne\the\@tempcnta\@citea\the\@tempcntb}\fi\fi}
\makeatother
\newenvironment{Eqnarray}%
     {\arraycolsep 0.14em\begin{eqnarray}}{\end{eqnarray}}

\def\be{\begin{equation}}
\def\ee{\end{equation}}
\def\bear{\be\begin{array}}
\def\eear{\end{array}\ee}
\def\bea{\begin{Eqnarray}}
\def\eea{\end{Eqnarray}}

\def\lsim{\mathrel{\raise.3ex\hbox{$<$\kern-.75em\lower1ex\hbox{$\sim$}}}}
\def\gsim{\mathrel{\raise.3ex\hbox{$>$\kern-.75em\lower1ex\hbox{$\sim$}}}}
\def\ifmath#1{\relax\ifmmode #1\else $#1$\fi}
\def\ls#1{\ifmath{_{\lower1.5pt\hbox{$\scriptstyle #1$}}}}

\def\beq{\begin{equation}}
\def\eeq{\end{equation}}
\def\beqa{\begin{Eqnarray}}
\def\eeqa{\end{Eqnarray}}

\def\gappeq{\mathrel{\rlap {\raise.5ex\hbox{$>$}}
{\lower.5ex\hbox{$\sim$}}}}
\def\lappeq{\mathrel{\rlap{\raise.5ex\hbox{$<$}}
{\lower.5ex\hbox{$\sim$}}}}

\begin{document}
%%%%%%%%%%%%%%%%%%%%%%%%%%%%%%%%%%%%%%%%%%%%%%%%%%%%%%%%%%%%%
\def\IJMPA #1 #2 #3 {{\sl Int.~J.~Mod.~Phys.}~{\bf A#1}\ (19#2) #3$\,$}
\def\MPLA #1 #2 #3 {{\sl Mod.~Phys.~Lett.}~{\bf A#1}\ (19#2) #3$\,$}
\def\NPB #1 #2 #3 {{\sl Nucl.~Phys.}~{\bf B#1}\ (19#2) #3$\,$}
\def\PLB #1 #2 #3 {{\sl Phys.~Lett.}~{\bf B#1}\ (19#2) #3$\,$}
\def\PR #1 #2 #3 {{\sl Phys.~Rep.}~{\bf#1}\ (19#2) #3$\,$}
\def\JHEP #1 #2 #3 {{\sl JHEP}~{\bf #1}~(19#2)~#3$\,$}
\def\PRD #1 #2 #3 {{\sl Phys.~Rev.}~{\bf D#1}\ (19#2) #3$\,$}
\def\PTP #1 #2 #3 {{\sl Prog.~Theor.~Phys.}~{\bf #1}\ (19#2) #3$\,$}
\def\PRL #1 #2 #3 {{\sl Phys.~Rev.~Lett.}~{\bf#1}\ (19#2) #3$\,$}
\def\RMP #1 #2 #3 {{\sl Rev.~Mod.~Phys.}~{\bf#1}\ (19#2) #3$\,$}
\def\ZPC #1 #2 #3 {{\sl Z.~Phys.}~{\bf C#1}\ (19#2) #3$\,$}
\def\PPNP#1 #2 #3 {{\sl Prog. Part. Nucl. Phys. }{\bf #1} (#2) #3$\,$}
%%%%%%%%%%%%%%%%%%%%%%%%%%%%%%%%%%%%%%%%%%%%%%%%%%%%%%%%%

%%%%%%%%%%%%%%%%%%%%%%%%%%% subequations.sty %%%%%%%%%%%%%%%%%%%%%%%%
\catcode`@=11
\newtoks\@stequation
\def\subequations{\refstepcounter{equation}%
\edef\@savedequation{\the\c@equation}%
  \@stequation=\expandafter{\theequation}%   %only want \theequation
  \edef\@savedtheequation{\the\@stequation}% % expanded once
  \edef\oldtheequation{\theequation}%
  \setcounter{equation}{0}%
  \def\theequation{\oldtheequation\alph{equation}}}
\def\endsubequations{\setcounter{equation}{\@savedequation}%
  \@stequation=\expandafter{\@savedtheequation}%
  \edef\theequation{\the\@stequation}\global\@ignoretrue

\noindent}
\catcode`@=12
%%%%%%%%%%%%%%%%%%%%%%%%%%%%%%%%%%%%%%%%%%%%%%%%%%%%%%%%%%%%%%%%%%%%%

\vspace*{-1in}
\renewcommand{\thefootnote}{\fnsymbol{footnote}}
\begin{flushright}
OUTP-01-50P \\
\end{flushright}
\vskip 5pt
\begin{center}
{\Large {\bf Charge and Color Breaking and D-terms \\
 in String Theory }}
\vskip 25pt
{\bf  Alejandro Ibarra \footnote{E-mail address:
ibarra@thphys.ox.ac.uk} }
 
\vskip 10pt  
{\it Department of Physics, Theoretical Physics, University of Oxford \\
 1 Keble Road, Oxford, OX1 3NP, United Kingdom}\\
\vskip 20pt
{\bf Abstract}
\end{center}
\begin{quotation}
  {\noindent\small

In four dimensional superstring models, 
the gauge group normally contains extra $U(1)$s
that are broken at a high energy scale. 
We show that the presence of the extra $U(1)$s is
crucial for the phenomenological viability of
string scenarios,
since the contribution to the scalar masses from the D-terms
 can lift the unbounded from below directions that 
usually appear when the gauge group at 
high energies is just the Standard Model
group. In particular, we show that the dilaton dominated scenario
can be allowed in large regions of the parameter space if there exists
an anomalous $U(1)$ in the theory, and the charges of the particles
are the appropriate ones.  
We also analyze the parameter space of
some explicit string constructions, imposing
a correct phenomenology and the absence of dangerous 
charge and color breaking minima or
unbounded from below directions.

\vskip 10pt
\noindent
}

\end{quotation}

\vskip 20pt  

\setcounter{footnote}{0}
\renewcommand{\thefootnote}{\arabic{footnote}}

\section{Introduction}

In a supersymmetric (SUSY) theory every fermion
has a corresponding scalar partner. As a result, the effective
potential is much richer than in a non-supersymmetric theory, 
and usually there are more minima apart from the electroweak (physical)
vacuum in which we live. The requirement that the electroweak vacuum
is the deepest minimum of the effective potential imposes very
strong constraints on the parameter space
\cite{CCB1,CCB2,tocho,nosotros,dilaton,cosmology}. For example, large
regions of the Constrained MSSM parameter space are excluded by the
requirement of the absence of charge 
and color breaking (CCB) minima and/or 
unbounded from below (UFB) directions \cite{tocho}.

In this paper, we would like to concentrate on four-dimensional
 string models, which have definite predictions on the soft
SUSY breaking terms. In these class of models, there are some massless
chiral superfields whose auxiliary components can acquire vacuum
expectation values (VEVs), and therefore break supersymmetry. These are the 
dilaton ($S$) and the moduli ($T_i$) fields, although some other fields 
could also contribute in some particular models. 
 There are several proposals
to trigger SUSY breaking in a hidden sector, leading to different 
soft terms. However, we prefer to follow 
\cite{Ibanez:1992hc,kaplu,brignole} and we will
assume that the underlying theory is a four dimensional string, and
that the fields that dominate the SUSY breaking are the dilaton and
the moduli. With these minimal assumptions, it is possible  to
parametrize the soft terms just with the dilaton and moduli auxiliary
fields ($F_S$, $F_{T_i}$), and thus obtain 
low energy predictions without addressing 
the problem of how is supersymmetry broken. A particularly interesting
limit in this parametrization corresponds to the case in which the 
dilaton is the field that dominates SUSY breaking. The dilaton field,
whose VEV determines the tree-level gauge coupling, is present in
any four-dimensional string model and couples at tree-level universally
to all particles. As a consequence, the soft terms are universal and 
independent of the four dimensional string considered. 
Furthermore, the  universality of the soft 
terms guarantees the absence of large 
flavour changing neutral currents and large CP effects.

A common feature of the string 
scenarios studied in the literature is that only in some small windows
our electroweak vacuum is the deepest minimum of the effective 
potential. Generically, the global 
minimum lies in a direction in which the stau acquires a VEV, 
thus breaking charge \cite{nosotros}. The case of
the dilaton dominated SUSY breaking limit  is especially 
disappointing, since {\it the whole} parameter space has 
a charge breaking global minimum \cite{dilaton}.
This is certainly a blow against these string scenarios, and some
possible ways-out have been proposed. It is possible that 
we  live in a metastable minimum, as long as its life time
is larger than the age of the Universe \cite{cosmology}. 
However, even it this is
the case, it is difficult to explain why the cosmological
evolution drove the Universe to a false vacuum instead of the true
vacuum. A second possibility assumes that the fundamental
theory is strongly coupled \cite{Witten:1996mz,Lykken:1996fj}. 
In that case, the
fundamental scale can be lowered and the CCB/UFB bounds on the dilaton
dominated scenario are relaxed, thus rescuing some regions of the 
parameter space \cite{Abel:2000bj}.

Many of these analyses assume that only the dilaton and moduli
auxiliary fields contribute to the soft terms.
However, most string constructions predict
the existence of extra gauge groups broken at
 a high energy scale that can contribute
to the scalar masses via D-term condensation \cite{Drees:1986vd}.
The goal of this paper is to study the impact of this contribution
on the CCB and UFB bounds in string scenarios. 

Let us briefly review the basic ingredients for our analysis. 
The general form of the soft SUSY-breaking Lagrangian in the MSSM 
is given by
\bea
\label{lsoft}
{\cal L}_{soft}=\frac{1}{2} \sum_{a=1}^{3}
M_a {\overline{\lambda}}_a \lambda_a  - \sum_{i} m_i^2 |\phi_i|^2
-(A_{ijk} W_{ijk} + B \mu H_1 H_2 + {\rm h.c.}),
\eea
where $W_{ijk}$ are the usual terms of the superpotential of the MSSM 
with $i=Q_L$, $u_R$, $d_R$, $L_L$, $e_R$, $H_1$, $H_2$, 
and $\phi_i,\lambda_a$ are 
the canonically normalized scalar and gaugino fields, respectively. 
On the other hand, $m_i, M_a$ and $A_{ijk}$ are the scalar masses, the
gaugino masses and the trilinear terms, respectively. String 
scenarios give definite predictions for these soft terms. However,
this is not the case for the bilinear term $B$, which depends crucially
on the mechanism that generates the $\mu$ term in the superpotential
($\mu H_1 H_2$). Although there are several interesting proposals 
to explain the origin of the $\mu$ term, the actual mechanism is still
unknown. 
Therefore, we  prefer to leave $B$ as a free parameter in
our analysis. As usual, the value of $\mu$ will be fixed by the
requirement of a correct electroweak symmetry breaking, i.e. by imposing
the existence of a realistic minimum with 
the correct mass for the $Z$ boson.

Now, we turn to the issue of the CCB and UFB bounds, discussed in 
detail in \cite{tocho}. There, it was found that the strongest constraint 
corresponds to the direction labeled as UFB-3, which
involves the fields $\{H_2,\nu_{L_i},e_{L_j},e_{R_j}\}$, with $i \neq j$,
and thus leads to electric charge breaking. For conciseness we will only 
write explicitly this one, since it is the most relevant one 
in our analysis. By simple analytical 
minimization of the relevant terms of the scalar potential, it is possible
to determine the value of the  ${\nu_{L_i},e_{L_j},e_{R_j}}$ fields in terms
of the $H_2$ one \cite{tocho}. 
Then, for any value of $|H_2|<M_{string}$ satisfying
\be
\label{SU6}
|H_2| > \sqrt{ \frac{\mu^2}{4\lambda_{e_j}^2}
+ \frac{4m_{L_i}^2}{g'^2+g_2^2}}-\frac{|\mu|}{2\lambda_{e_j}} \ ,
\ee
the value of the potential along the UFB-3 direction is simply given
by
\be
\label{SU8}
V_{\rm UFB-3}=(m_2^2 -\mu^2+ m_{L_i}^2 )|H_2|^2
+ \frac{|\mu|}{\lambda_{e_j}} ( m_{L_j}^2+m_{e_j}^2+m_{L_i}^2 ) |H_2|
-\frac{2m_{L_i}^4}{g'^2+g_2^2} \ .
\ee
Otherwise
\be
\label{SU9}
V_{\rm UFB-3}= (m_2^2 -\mu^2 ) |H_2|^2
+ \frac{|\mu|} {\lambda_{e_j}} ( m_{L_j}^2+m_{e_j}^2 ) |H_2| + \frac{1}{8}
(g'^2+g_2^2)\left[ |H_2|^2+\frac{|\mu|}{\lambda_{e_j}}|H_2|\right]^2 \ .
\ee
In eqs.(\ref{SU8},\ref{SU9}) $\lambda_{e_j}$ is the leptonic Yukawa
coupling of the $j-$generation and $m_2^2$ is the sum of the $H_2$ 
soft mass squared, $m_{H_2}^2$, plus $\mu^2$. Then, the
UFB-3 condition reads
\be
\label{SU7}
V_{\rm UFB-3}(Q=\hat Q) > V_{\rm real \; min} \ ,
\ee
where $V_{\rm real \; min}=-\frac{1}{8}\left(g'^2 + g_2^2\right)
\left(v_2^2-v_1^2\right)^2$, with $v_{1,2}$ the VEVs of the Higgses $H_{1,2}$,
is the realistic minimum evaluated at $M_S$ (see below)
and the $\hat Q$ scale is given by \linebreak
$\hat Q\sim {\rm Max}(g_2 |e|, \lambda_{top} |H_2|,
g_2 |H_2|, g_2 |L_i|, M_S)$
with
$|e|$=$\sqrt{\frac{|\mu|}{\lambda_{e_j}}|H_2|}$ and
$|L_i|^2$=$-\frac{4m_{L_i}^2}{g'^2+g_2^2}$ \linebreak +($|H_2|^2$+$|e|^2$).
Finally, $M_S$ is the typical scale of SUSY masses (normally a good
choice for $M_S$ is an average of the stop masses).
Notice from (\ref{SU8},\ref{SU9}) that 
the negative contribution to $V_{UFB-3}$
is essentially given by the $m_2^2-\mu^2$ term, which can be very sizeable in 
many instances. On the other hand, the positive contribution is dominated by 
the term $\propto 1/\lambda_{e_j}$, thus the larger
$\lambda_{e_j}$ the more restrictive
the constraint becomes. Consequently, the optimum choice of
the $e$--type slepton is the third generation one, i.e.
${e_j}=$ stau.

It is apparent from eq. (\ref{SU8}) that the UFB-3 bound is relaxed 
when the slepton masses are large compared to the
rest of the parameters in the scalar 
potential  \cite{nosotros}. There is also 
a second order effect that relaxes
the UFB-3 bound, namely when the squark masses are small. 
The reason is that $m^2_{H_2}$ is driven negative by the stop
contribution to its RGE, $Q dm^2_{H_2}/dQ=6(\lambda_{top}/4 \pi)^2
(m^2_{\tilde t_L}+m^2_{\tilde t_R})+...$,
hence the smaller $m^2_{\tilde t_L}$,
$m^2_{\tilde t_R}$ the weaker the UFB-3 bound becomes 
\footnote {Also, a departure of gaugino universality in the
direction $M^2_2,M^2_1>M^2_3$ helps to rescue regions in the parameter
space. Incidentally, this is the pattern expected in
weakly coupled  string scenarios, when  the one loop
corrections to the gauge kinetic functions 
are included. Nevertheless, throughout
the paper we will consider universal gaugino masses, neglecting
the above-mentioned effect.}. Some of the superstring 
scenarios analyzed in \cite{nosotros} did not
have this kind of spectrum, and therefore the UFB-3 bound was
devastating. However, the D-term contribution to the scalar
masses, which appears in most explicit superstring constructions
when some extra gauge groups are broken at a high energy scale,
can modify the spectrum of scalar masses to produce one
 with this characteristics. If this is the case,
large regions of the parameter space can be rescued. 

The paper is organized as follows. In Sect. 2 we analyze the
D-term contribution to the scalar masses 
in a superstring inspired scenario with an
anomalous $U(1)$, and examine the implications for
CCB minima and UFB directions. In Sect. 3 we perform a 
similar analysis for the case of a non-anomalous $U(1)$.
Sect. 4 is devoted to the analysis of two realistic 
superstring scenarios, with several extra $U(1)$s. 
Finally, the conclusions are presented in  Sect. 5.

\section{Scalar masses in the presence of an anomalous $U(1)$}

In many explicit four-dimensional heterotic string 
constructions, the massless spectrum includes matter chiral 
superfields that transform under an anomalous 
$U(1)$ symmetry according
to $\phi_k \rightarrow e^{i q_k \Lambda} \phi_k$, and
a vector superfield that transforms as 
$V_A \rightarrow V_A + i (\Lambda - \Lambda^*)$,
where $\Lambda$ is the transformation parameter. 
The anomalies are compensated by a non-trivial transformation
of the dilaton under the  $U(1)_A$, 
$S \rightarrow S+ i \Lambda \delta_{GS}$, according to
the Green-Schwarz mechanism \cite{Green:1984sg}. In a string theory, 
the coefficient $\delta_{GS}$ can be computed explicitly and turns out
to be proportional to the apparent chiral anomaly \cite{deltaGS}
\be
\label{deltaGS}
\delta_{GS}= {1 \over {192 \pi^2}} \sum _k q_k.
\ee

The charges of the fields under the anomalous $U(1)$ are
severely constrained by the Green-Schwarz mechanism.
The anomalies are cancelled only if 
the coefficients $A_i$ of the mixed anomalies of
the $U(1)_A$ satisfy the condition $A_i/k_i=\delta_{GS}$,
where $k_i$ is the Kac-Moody level of the gauge factor.
In particular, the mixed anomalies with the Standard Model groups,
 $SU(3)$, $SU(2)$ and $U(1)_Y$, must be in 
the ratio $A_3:A_2:A_1=k_3:k_2:k_1$.
 Normally (one level case) one takes
$k_3=k_2=\frac{3}{5}k_1=1$ to account for the unification of the
gauge couplings, which translates into $A_3=A_2=\frac{3}{5} A_1=1$
\cite{Ibanez:1993fy}. If the only fields charged under the SM gauge
group are the usual fields of the MSSM, 
it can be proved that there are only two family 
independent $U(1)$ symmetries  \footnote{We
restrict ourselves to family independent symmetries for reasons
that will be explained later.} for which $A_3=A_2=\frac{3}{5} A_1=1$.
Following ref. \cite{Ibanez:1994ig}, we
will denote them by  $U(1)_X$ and $U(1)_{XX}$. The most
general ``anomalous'' $U(1)$ symmetry is a combination
of those two, plus all the possible traceless $U(1)$, 
since they do not
modify the conditions for anomaly cancellation.
 With the minimal particle content of the MSSM, 
there is only one such symmetry (apart from
the weak hypercharge) which we denote by $U(1)_H$.
The charges of the MSSM fields under those groups are 
\be
\label{symmetries}
\begin{tabular}{c|ccccccc}
   & $Q_L$ & $ u_R$ & ${d_R}$ &
${L_L}$ & ${e_R}$ & ${H_2}$ & $ {H_1}$  \\
\hline
  $U(1)_{H}$ & 0 & 0 & 0 & 0 & 0 & 1 & --1
\\
  $U(1)_{X}$ & 1 & 1 & 0 & 0 & 1 & 0 & 0
\\
  $U(1)_{XX}$ & 0 & 0 & 1 & 1 & 0 &  0&  0
\\
\end{tabular}
\ee
Then, the most general ``anomalous'' $U(1)$, with  anomalies
cancelled by the Green-Schwarz mechanism, is a linear combination
of those symmetries:
\be
\label{U1}
U(1)_A= z \, U(1)_H +x \,  U(1)_X +y \, U(1)_{XX}.
\ee

Now, we turn to describe the corresponding supergravity theory.
It is defined by three functions, namely
the K\"ahler potential $K(\phi_k,\phi^*_k)$, the gauge
kinetic function $f_a(\phi_k)$ and the superpotential $W(\phi_k)$.
Gauge invariance requires the K\"ahler potential to be a function
of $S+S^*-\delta_{GS} V_A$. 
In particular, for orbifold compactifications, the 
tree-level K\"ahler potential reads \cite{kahler}
\be
\label{kahler1}
K=-\log (S+S^*-\delta_{GS} V_A)-3 \log(T+T^*)+
(T+T^*)^{n_{\phi_k}} e^{2 q_k V_A}  \phi_k^* \phi_k,
\ee
where ${n_{\phi_k}}$ is the modular weight of the field $\phi_k$.
We assume that in  the massless spectrum there is one
chiral superfield, denoted by $\Phi$, with charge $Q$ under
$U(1)_A$ and singlet with respect to the Standard Model group,
that will eventually acquire a VEV,
thus breaking the anomalous symmetry.
On the other hand, the gauge kinetic function is, at tree
level, $f_a=k_a S$, where $k_a$ were defined above.
Finally, let us discuss the superpotential in this class of models.
Commonly, the terms that appear in the MSSM superpotential, that we
denote generically by $\phi_i \phi_j \phi_k$, are not 
gauge invariant under $U(1)_A$. Instead, the superpotential reads
\be
\label{superpotential}
W \sim \Theta [-(q_i+q_j+q_k)/Q]
\left( \frac{\Phi}{M_{Pl}} \right) ^{-(q_i+q_j+q_k)/Q}
 \phi_i \phi_j \phi_k,
\ee
where $\Theta(x)$ is a step function, defined as 1 when $x \geq 0$
and 0 otherwise, to account for the requirement of holomorphicity.
Then, if we want to end up at low energies with
the MSSM superpotential, the charges of the particles under $U(1)_A$
must satisfy 
\bea
\label{holomorphicity}
&(q_{Q_L}+q_{u_R}+q_{H_2})/Q \leq 0& \;\; ,\nonumber\\ 
&(q_{Q_L}+q_{d_R}+q_{H_1})/Q \leq 0& \;\; ,\nonumber\\
&(q_{L_L}+q_{e_R}+q_{H_1})/Q \leq 0& \;\; .
\eea
These impose further
constraints on the charges of the particles under the anomalous $U(1)$,
apart from the condition of anomaly cancellation.
A similar holomorphicity constraint applies
to the bilinear term, $(q_{H_1}+q_{H_2})/Q \leq 0$.

In the low energy theory, the scalar particles receive two
different contributions to  their masses. The first one 
depends on  which are the
fields responsible for the SUSY breaking, 
$S$ and/or $T$. These have been computed
in \cite{brignole} in terms of the goldstino angle, defined as
$\tan \theta= F_S/F_T$, and the
gravitino mass $m_{3/2}$. 
Assuming cancellation of the cosmological constant, one obtains 
from eq.(\ref{kahler1}) a contribution to the scalar masses given by
\be
\label{m-F}
[m^2_k]_F=m_{3/2}^2(1+n_k \cos^2 \theta).
\ee

The second contribution to the scalar masses comes from the 
breaking of the $U(1)_A$ symmetry, and contributes through the
expectation value of the D-term associated to $U(1)_A$. 
The anomalous $U(1)$ generates a Fayet-Iliopoulos (F-I) D-term
in the scalar potential that can break supersymmetry \cite{FI}:
\be
\label{D-term}
D_A=-g_A^2(\xi^2+q_k |\phi_k|^2),
\ee
where $\xi^2=-\delta_{GS}/2(S+S^*)$ is the F-I term. Large-scale
supersymmetry breaking is avoided through the chiral field $\Phi$
acquiring  a vacuum expectation
value that nearly cancels the F-I term. Due to the low-scale
supersymmetry breaking, the cancellation is not
exact, and this amounts to a contribution to the scalar masses, 
given by

\be
\label{m-D}
[m^2_k]_D=-q_k \langle D_A \rangle.
\ee

Assuming that $U(1)_A$ is broken 
at a scale much smaller than the gravitational scale,
 but much larger than the electroweak scale \footnote
{For an anomalous $U(1)$, the scale of symmetry breaking can
be read from the condition of D-flatness. It yields 
$ \langle \Phi \rangle \sim 0.02 \sqrt{\sum_k q_k/Q} M_{Pl}$,
that typically satisfies these conditions.}, one obtains 
\cite{Kawamura:1996wn,Kawamura:1997bd}
\be
\label{vev-D-a}
\langle D_A \rangle= -\frac {1} {Q} m_{3/2}^2 
(1-6 \sin^2 \theta - n_{\Phi} \cos^2 \theta).
\ee

At this point, some comments are in order. First, notice that 
the positivity of the scalar masses is not guaranteed
after the breaking of $U(1)_A$. This could lead to
the breaking of some of the Standard Model symmetries at
a scale $\sim \xi$, including charge and/or color. This
in turn imposes constraints on the goldstino angle and the 
charges of the particles under $U(1)_A$. Secondly, if the
anomalous $U(1)$ symmetry is horizontal, 
the family-dependent contributions to the scalar 
masses from the D-terms produce rates
for the flavour changing processes that could be 
sizeable. Therefore, to avoid potential problems with flavour violation, 
we will assume that the anomalous $U(1)$ symmetry 
is family independent.

The gaugino masses and the trilinear soft terms
do not receive any contribution from the D-terms.  At tree 
level in the gauge kinetic function and the K\"ahler potential, and
neglecting phases, they are given by \cite{brignole}
\be
\label{gaugino}
M=\sqrt{3} m_{3/2} \sin \theta \;\; ,
\ee
\be
\label{trilinear1}
A_{ijk}=-\sqrt{3}\; m_{3/2} 
[\sin \theta+\cos \theta \omega_{ijk}(T,T^*)]  \;\; ,
\ee
where
\be
\label{omega}
\omega_{ijk}(T,T^*)=\frac 1 {\sqrt{3}}\left (3+n_i+n_j+n_k-(T+T^*)
\frac{\partial \lambda_{ijk}/\partial T}{\lambda_{ijk}} \right) \;\; ,
\ee
and $\lambda_{ijk}$ are the Yukawa couplings in the superpotential.

Next, we study the impact of the anomalous $U(1)$ symmetry 
on different well motivated four-dimensional string scenarios.

\subsection{The dilaton limit}

The dilaton limit is a very attractive scenario, 
since it alleviates the flavour and the CP
problems in supersymmetric theories.
In this scenario, the different terms in the soft SUSY-breaking
Lagrangian have a particularly simple form:
\bea
\label{softterms}
m_k^2 &=& m_{3/2}^2 (1+5 q_k)\;\; ,\nonumber\\ 
M &=& \pm \sqrt{3} \; m_{3/2} \;\;,\nonumber\\
A &=& -M \;\;.
\eea
where we have assumed, without loss of generality, that $Q=-1$.

In the Introduction, we remarked that the UFB-3 bound is relaxed
when the slepton masses are large and the squark masses are small, 
the former being the dominant effect. It is interesting 
to note that the sleptons can be charged under the most
general anomaly-free $U(1)$, eq.(\ref{U1}),
 and their masses can receive
a contribution from the D-terms that could lift the UFB-3
direction. To decide if that is possible or not,  we study first
some particular choices for $U(1)_A$ and later on, we will discuss the
most general case.

The assignment of charges for the $U(1)_{XX}$ symmetry yields 
$m^2_{d_R}=m^2_{L_L}= 6 m_{3/2}^2$, 
while the rest of the masses remain unchanged. 
Also, with this assignment of charges, the requirement of
holomorphicity, eqs. (\ref{holomorphicity}), is
satisfied, and we will get at
low energies all the Yukawa couplings of the Standard Model
Lagrangian.
We plot the regions of the parameter space forbidden
from the different UFB and CCB bounds, as well as from the
condition of a correct top mass and the experimental lower bounds
on the supersymmetric masses. Here 
we choose the positive sign for the gaugino mass, although the
results for negative gaugino masses can also be read from the
figures, due to the invariance of the analysis under the
transformation $B,A,M \rightarrow -B,-A,-M$.
Then, the only free parameters are the gravitino mass ($m_{3/2}$)
and the bilinear soft term ($B$), so,  we will use these variables
to span the parameter space.
In Figure 1,  we show the results
assuming that the only contribution to the scalar masses comes
from the dilaton and moduli auxiliary fields 
(or equivalently, when the gauge group at
high energies is just the Standard Model group, without an extra
anomalous $U(1)$ in the theory). This was the case studied
 in \cite{dilaton}, where it was found that the whole
parameter space is excluded. On the other hand, 
in Figure 2, left plot,
 we show the results for the assignment of charges corresponding to
$U(1)_{XX}$. 
All the points that were forbidden 
by the UFB-3 bound now become allowed.
The reason is that $m^2_{L_L}$ is large enough to compensate
the negative contribution from $m^2_2-\mu^2$ in eq.(\ref{SU8}).
Also, $m^2_{L_L}$ appears in the coefficient linear with $|H_2|$,
which is positive, and this further alleviates the UFB-3 
constraint. However, this effect is not as
strong as the above-mentioned one.
 Notice also that there are
no points forbidden by CCB bounds. The bound that was
not satisfied in Figure 1 was
\be
\label{CCB}
|A_{\tau}|^2 \leq 3(m^2_{L_L}+m^2_{e_R}+m^2_{H_1}),
\ee
which is obviously relaxed when the slepton masses are large. 
This example illustrates the tremendous impact 
that the D-term contribution to the scalar masses
can have on the low energy phenomenology.

\begin{figure}
\begin{center}
\epsfig{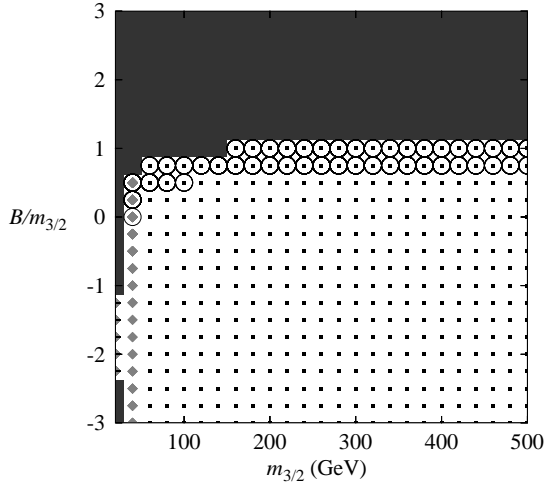}
\caption{Excluded regions of the parameter space assuming SUSY 
breaking dominated by the dilaton, in a scenario where the
gauge group at high energies is just 
the Standard Model gauge group.
The black regions are excluded because it is not possible to
reproduce the experimental mass of the top quark. The small squares
indicate regions excluded by unbounded from below bounds, 
while the circles, regions excluded by charge and 
color breaking constraints. Finally, the filled diamonds correspond to
regions excluded by the experimental lower bounds on supersymmetric
masses.}
\end{center}
\end{figure}

On the other hand, the charges for $U(1)_{X}$ give rise to a 
different spectrum of scalar masses,
namely,
$m^2_{Q_L}=m^2_{u_R}=m^2_{e_R}=6 m_{3/2}^2$  and the rest of the
scalar masses equal to $m^2_{3/2}$. This spectrum also has 
a heavy slepton, who helps to lift the UFB-3 direction. Unfortunately,
$m^2_{e_R}$ only enters in the term linear with $|H_2|$ in 
eqs.(\ref{SU8},\ref{SU9}), and is not enough to rescue points in the
parameter space. A large $\tilde{e}_R$ mass is useful,
though, to rescue points forbidden by
the CCB constraint eq.(\ref{CCB}), but still the whole
parameter space is excluded by UFB constraints. The numerical
results are shown in Figure 2, right plot.

\begin{figure}
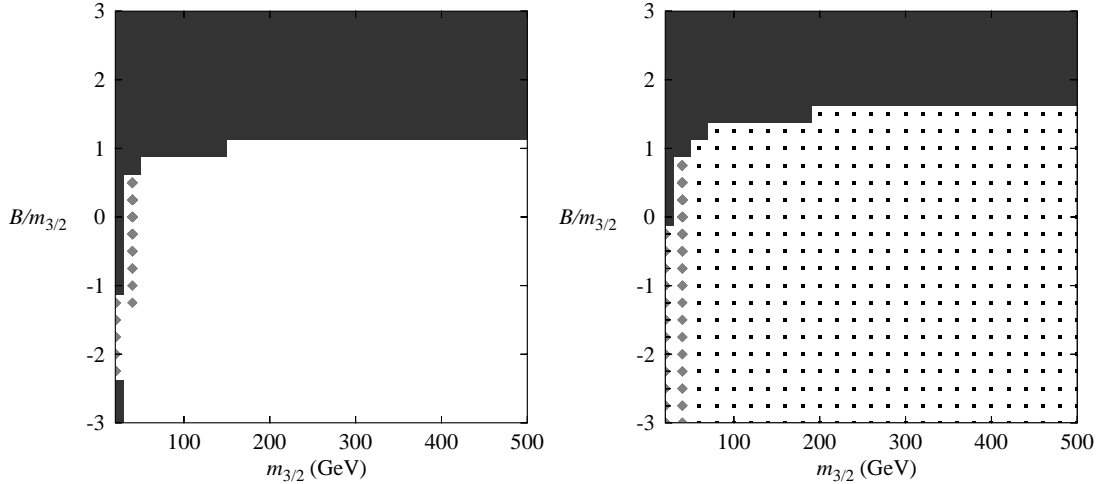

\begin{center}
\epsfig{file=fig2a.ps,height=6.5truecm}
\epsfig{file=fig2b.ps,height=6.5 truecm}
\caption{The same as Figure 1, but extending the Standard Model 
gauge group with
the anomalous symmetry $U(1)_{XX}$ (left) or $U(1)_{X}$ (right).
The meaning of the symbols is the same as in Figure 1.}
\end{center}
\end{figure}

The remaining symmetry is $U(1)_H$. We do not analyze this symmetry
separately because by itself is  not able to implement
the Green-Schwarz mechanism (all the mixed anomalies with the SM
gauge group are vanishing). Also, the fact that at least one
Higgs has negative charge under $U(1)_H$ implies that the corresponding
scalar mass squared can be negative, unless $|z| \leq 1/5$. 
In terms of charges, this
translates into $|q_{H_2}/Q| \leq 1/5$, 
which does not  seem very natural. 
We should stress that the presence of a 
negative Higgs mass squared does not 
represent a problem for charge breaking, since in the MSSM the
minimum of the Higgs potential always lies at  
$\langle H^+_2 \rangle=\langle H^-_1 \rangle=0$. It could
be a problem, though, for the electroweak symmetry
breaking. A correct electroweak symmetry breaking
imposes the following constraint on the parameters entering
in the Higgs potential:
\be
\label{UFB-1}
m^2_{H_1}+m^2_{H_2}+2 \mu^2 - 2 | \mu B|>0.
\ee
Then, if the scalar masses are negative already at the high
energy scale, this bound is more difficult to fulfill. 
However, it should be noted that it could
be possible to find some regions in the parameter space where
the bound eq.(\ref{UFB-1}) is satisfied, and the electroweak
symmetry is broken in the correct way.
For simplicity, we will not consider this possibility here and we will
impose that at high energies all the scalar masses are positive. 
So, a sensible choice would be to assume $z=0$, although our
conclusions for this scenario are not very different if 
 $|z| \leq 1/5$.

Then, the most general anomaly-free $U(1)$ is just a combination
of $U(1)_X$ and $U(1)_{XX}$. 
%The requirements of holomorphicity and 
%positivity of the scalar masses are fulfilled if $x,y \ge 0$.
We argued that the UFB-3 direction was lifted when the $\tilde{L}_L$
mass was large compared to the rest of the scalar masses, especially
 $m^2_{H_2}$. This was achieved by the charge assignment of the
$U(1)_{XX}$ symmetry, whereas the $U(1)_X$ symmetry did not have any
significant effect on the UFB bounds.
Then, we expect that if $U(1)_{XX}$ is
present in $U(1)_A$ to some extent,
some regions of the parameter space could
be rescued, as long as the
requirements of positivity and holomorphicity are fulfilled. To 
show this, we study 
\be
\label{U1_2}
U(1)_A=  x U(1)_X +(1-x) U(1)_{XX},
\ee
and we plot the results for a fixed gravitino mass ($m_{3/2}=100$ GeV).
Taking $x=0$ corresponds to the limiting case 
in which the anomalous symmetry
is just $U(1)_{XX}$ (no points forbidden by UFB or CCB bounds),
while $x=1$ corresponds to $U(1)_{X}$ (the whole parameter space is 
excluded). If $x>1$ or $x<0$ positivity is not satisfied.
We find that all our phenomenological requirements are satisfied
for $x \lsim 0.4$, as is shown in Figure 3.
If the gravitino mass is larger, or the charges of all the MSSM fields
are larger [for example, if 
$U(1)_A=  2 \,x\, U(1)_X +2\,(1-x)\, U(1)_{XX}$],
the allowed region becomes larger as well.

\begin{figure}
\begin{center}
\epsfig{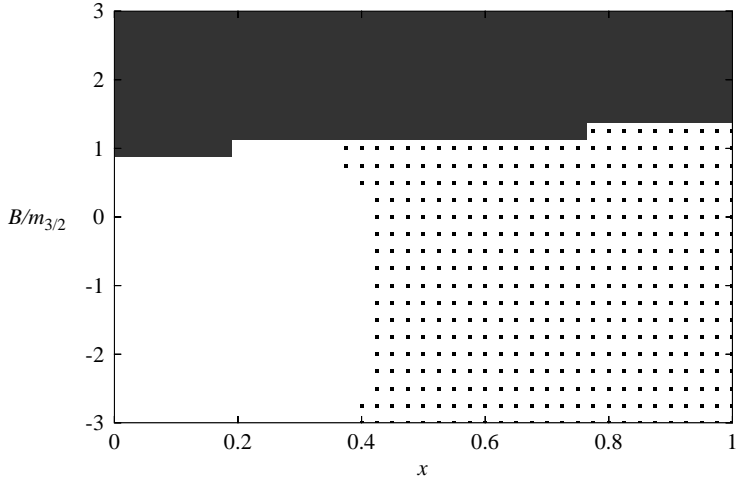}
\caption{The same as Figure 1, but extending the Standard Model 
gauge group with
the anomalous symmetry $U(1)_A=x U(1)_X +(1-x) U(1)_{XX}$, for a 
gravitino mass of 100 GeV.
The meaning of the symbols is the same as in Figure 1.}
\end{center}
\end{figure}

At this point, we should remind the reader
that we have analyzed a {\it string inspired} 
scenario, with just the MSSM content 
plus an additional superfield, $\Phi$, singlet under
the SM gauge group, to
break the anomalous $U(1)_A$. In realistic string models, there
are in general more fields, with masses of ${\cal O}(\xi)$, charged 
under the Standard Model symmetries, that modify the 
conditions for mixed anomaly cancellation.
Therefore, it could happen that the $U(1)_A$ symmetry
is not of the form eq.(\ref{U1}), and in consequence,
the analysis should be done case by case. Nevertheless, it is 
encouraging that the dilaton limit 
can be rescued by the presence of an anomalous $U(1)$.
We will study some explicit string constructions 
in Section 4.

\subsection{Modular weights equal to --1}

In this subsection we consider a scenario where all the relevant
particles have modular weights equal to --1. This is the case
when the observable sector belongs to the untwisted sector of an
orbifold compactification, although the results are also valid
for a Calabi-Yau compactification in the large-$T$ limit. 
For the purposes of this paper, this
case is interesting because it permits the study of the effect of the
goldstino angle on the region forbidden by the UFB-3 bound, after
including the D-term  contribution to the scalar masses. Note that
different modular weights for different particles would produce 
by itself a non-universality in the scalar masses that could obscure
the effect we want to study. For this scenario, the Yukawa couplings
$\lambda_{ijk}$ are T-independent since a cubic operator of matter
fields has exactly the appropriate modular weight (--3). Therefore, 
the expression for the trilinear term 
given in eq.(\ref{trilinear1}) becomes independent 
of $\lambda_{ijk}$ and are universal. Under the
usual assumptions, the different terms appearing
in the soft SUSY-breaking Lagrangian are:
\be
\label{escalar2}
m^2_k=m^2_{3/2}[\sin^2 \theta + 
\frac{q_k}{Q} (1-6 \sin^2 \theta +\cos^2 \theta)]   \;\; ,
\ee
\be
\label{gaugino2}
M=\sqrt{3} m_{3/2} \sin \theta \;\; ,
\ee
\be
\label{trilinear2}
A_{ijk}=-\sqrt{3} m_{3/2} \sin \theta \;\; .
\ee

If the $U(1)_A$ charges of the MSSM fields 
are all set to zero, we find
that these expressions satisfy the relation 
$M=-A=\pm \sqrt{3}m$,  independently of $\theta$. 
Since the CCB and UFB bounds depend, for a given $B$, only on $M/m$ and 
$A/m$, in every point of the parameter space the results
are  identical to the results at $\theta=0$, i.e.
the dilaton limit, which is excluded.

The presence of the anomalous $U(1)$ in this scenario
has several consequences. The first one  is a 
non-trivial dependence of the results with the goldstino angle,
as is evident from eq.(\ref{escalar2}) and the previous discussion.
Also,  in this scenario it is possible 
to have negative scalar
masses squared at the high energy scale, in contrast to the case
with $q_k=0$, where the positivity 
was guaranteed in every point. As was
mentioned in \cite{Kawamura:1996wn}, the condition for positivity is
\be
\label{positivity-1}
( 1-7 q_k/Q)\sin^2 \theta +2 q_k/Q \ge 0 \;.
\ee
It is interesting to note that in the moduli limit,  $\sin \theta = 0$, 
positivity requires $q_k/Q \ge 0$, which is in conflict
with the holomorphicity constraints, eqs.(\ref{holomorphicity}), 
except in the trivial case $q_k/Q=0$. 

Most of the conclusions drawn for the dilaton dominated scenario 
are also valid here: the existence of an anomalous group can rescue
large regions of the parameter space, 
when the charges are such that the $\tilde{L}_L$ mass 
is larger than the rest.
To illustrate this, we study in detail the assignment of charges
corresponding to the $U(1)_{XX}$ symmetry in eq.(\ref{symmetries}).
 Positivity of the masses squared at the high
energy scale requires $\sin \theta > 1/2$ (as before, we are
setting $Q=-1$), and accordingly we
present the results only for that
range of the goldstino angle. We have just plotted for
$\theta$ in the $[0,\pi]$ range because the results in the
$[\pi,2 \pi]$ range can also be read from the figures, due to
the invariance of the analysis under the
transformation $B,A,M \rightarrow -B,-A,-M$.
The regions allowed by all our
phenomenological requirements are shown in Figure 4. There are large
regions that have become allowed, 
while there are still some windows forbidden
by the UFB-3 bound. These correspond to  points where the 
non-degeneracy of the masses is not large enough to lift the 
UFB-3 direction. On the other hand,  the parameter
space for the $U(1)_X$ symmetry is entirely excluded, 
as expected from the results for the dilaton limit.
In a more general case, when the ``anomalous'' $U(1)$ symmetry
is a combination of $U(1)_X$ and $U(1)_{XX}$,
 as in eq. (\ref{U1}), there appear allowed regions for a 
wide range of goldstino angles  when $x \lsim 0.4$  
(or  may be more, depending on $m_{3/2}$
and the charges).

\begin{figure}
\begin{center}
\epsfig{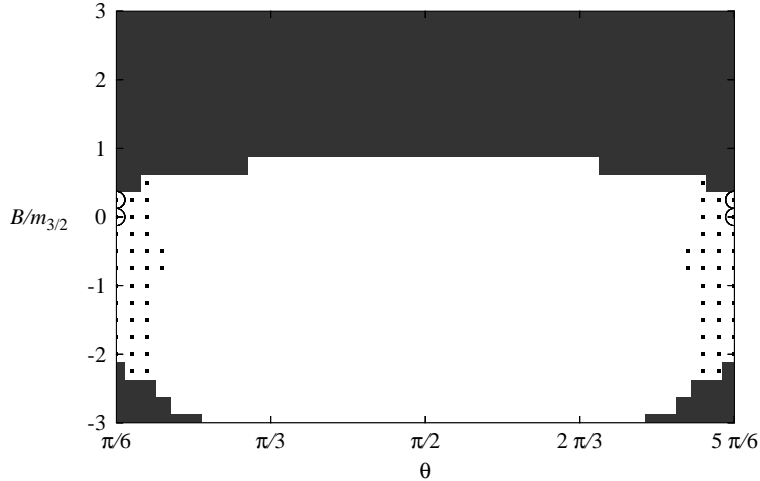}
\caption{Allowed regions for a scenario where all the matter fields
have modular weight $-1$ and the Standard Model 
gauge group has been extended
with the $U(1)_{XX}$ symmetry. We have fixed the gravitino mass to
100 GeV and the meaning of the symbols is the same as in Figure 1.}
\end{center}
\end{figure}

\subsection{ILR model}

This model was proposed by Iba\~nez, L\"ust 
and Ross \cite{Ibanez:1991zv}
and has the nice feature that it provides 
string threshold corrections large enough 
to fit the joining of the gauge
coupling constants at a scale $\simeq 10^{16}$ GeV. This is
achieved by assigning the following modular weights to 
the MSSM fields:
\begin{eqnarray}
n_{Q_L}=n_{d_R}=-1\ ,\ \ n_{u_R}=-2\ ,\ \
 n_{L_L}=n_{e_R}=-3 \ ,\ \ 
n_{H_1}+n_{H_2}=-5,-4 \ .
\label{F0010}
\end{eqnarray}
The above values together with a Re$T\simeq 16$ lead
to a good agreement for $\sin^2\theta _W$ and $\alpha _3$.
This case is also interesting because there are two 
sources of non-universality in the scalar masses: 
in the contribution from the moduli auxiliary field, due to
the different modular weights of the MSSM fields, and in
the  contribution from the D-term, due to the different charges 
under $U(1)_A$. 
The assignment of modular weights
makes the slepton soft masses smaller than the
rest, hence, if the gauge group at high energies was
just the SM gauge group,
the UFB-3 direction would be rather deep (in fact, in
that case, the whole parameter space is excluded \cite{nosotros}).
We want to study here if the D-term contribution to the scalar
masses can be large enough to lift the UFB-3 direction, 
even in this particularly unfavorable scenario.

As in the previous subsection, we extend the 
SM gauge group with the $U(1)_{XX}$ symmetry,
defined in eq.(\ref{symmetries}), and the matter content of the MSSM
with a field $\Phi$, only charged under  $U(1)_{XX}$. The
scalar masses can be readily computed from eqs.(\ref{m-F},
\ref{m-D}, \ref{vev-D-a}). On the other hand, the
computation of  the trilinear soft terms, 
eqs.(\ref{trilinear1},\ref{omega}), deserves
some more explanation. In this case, the Yukawa couplings, $\lambda_{ijk}$,
are non-trivial functions of $T$ \cite{Hamidi:1987vh}, which apparently
causes a difficulty when determining the trilinear soft terms.
However, the top Yukawa coupling, the only  relevant one for
the evaluation of the UFB-3 bound, represents a remarkable 
(and fortunate) exception.
The reason is the following. The twisted Yukawa couplings
are in general given by a series of terms, all of them suppressed
by $e^{-c_iT}$ factors. Only when the  fields involved belong
to the same fixed point (or fixed torus) the first term in the series
is $O(1)$ and independent of $T$, otherwise the coupling is suppressed.
Consequently, the top Yukawa coupling, being $O(1)$, must be of
this type. So, for this particular case we can ignore the 
$\partial \lambda/\partial T$ factors in (\ref{omega}),
thus getting, for the assignment of modular weights corresponding
to the ILR model,
$\omega= n_{H_2}/\sqrt{3}$, $|A|=- m_{3/2}(\sqrt{3} \sin \theta
+ n_{H_2} \cos \theta)$.

The assignment given in eq. (\ref{F0010}) allows different choices of
the modular weights of the Higgs doublets. We show the results
for the case $n_{H_1}=-2$ and $n_{H_2}=-3$,
 and $n_{\Phi_1}= n_{\Phi_2}=-1$, although the conclusions
for other choices are completely analogous. 
In Figure 5 we show the numerical
results for $\cos^2 \theta \leq 1/3$, to avoid negative
scalar masses squared at the high energy scale (in particular,
 $m^2_{e_R}$). The contribution to the scalar masses coming
from the D-terms succeeds to modify the spectrum of scalar masses
from one with light sleptons, unfavorable from the point
of view of the UFB-3 bound, to one with heavy sleptons.
In consequence, large regions of the parameter space become
allowed after the inclusion of the D-term contribution to the
scalar masses. In a more general case, in which the anomalous
gauge group is a combination of $U(1)_X$ and $U(1)_{XX}$,
as in eq.(\ref{U1_2}),
this scenario becomes allowed for $x \lsim 0.2$. As 
expected, it is more difficult to rescue this scenario than 
the one with $n=-1$, since at the gravitational scale, before
the breaking of the anomalous $U(1)$, the scalar masses are already
non-universal and have a spectrum which is rather unfavorable
from the point of view of the UFB bounds.

\begin{figure}
\begin{center}
\epsfig{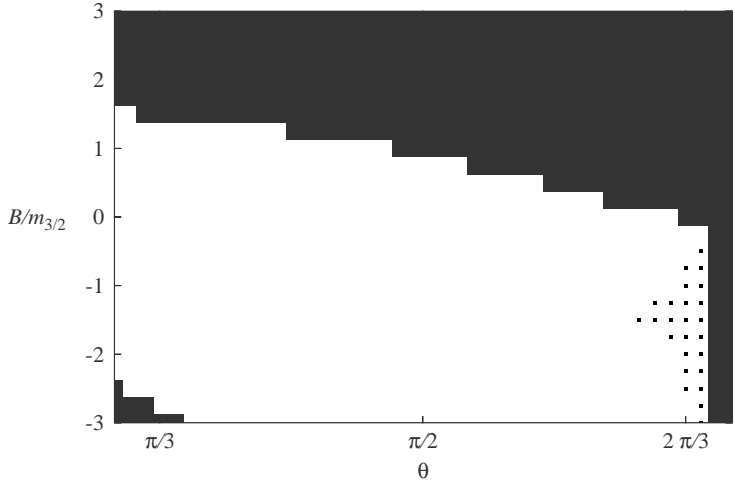}
\caption{
Allowed regions for the ILR scenario where the Standard Model 
gauge group has been extended
with the $U(1)_{XX}$ symmetry. The gravitino mass is fixed to
100 GeV and the meaning of the symbols is the same as in Figure 1.
}
\end{center}
\end{figure}

\section{Scalar masses in the presence of a non-anomalous $U(1)$}

The presence of non-anomalous $U(1)$s is also quite common in 
four-dimensional string constructions. Hence, it is worth studying
the effect of the breaking of these groups on the charge and color
breaking bounds.

Let us assume a simplified scenario in which there is only one
non-anomalous $U(1)$, whose gauge supermultiplet we denote by $V$. 
Then, for orbifold compactifications,
the tree level K\"ahler potential is
\be
\label{kahler2}
K=-\log (S+S^*)-3 \log(T+T^*)+
(T+T^*)^{n_{\phi_k}} e^{2 q_k V}  \phi_k^* \phi_k,
\ee
where $q_k$ and ${n_{\phi_k}}$ are, respectively, the $U(1)$ charge and 
the modular weight of the chiral superfield $\phi_k$. 
Let us also assume that in the massless spectrum there are
two chiral superfields, $\Phi_1$ and $\Phi_2$, with charges
$Q$ and $-Q$ and modular weights $n_{\Phi_1}$ and $n_{\Phi_2}$, 
respectively, whose scalar components acquire a vacuum
expectation value and break the $U(1)$ symmetry at a scale
 much smaller than the gravitational scale but much larger
than the electroweak scale. Under these assumptions, the
D-term contribution to the scalar masses is:

\be
\label{m-D-noanom}
[m^2_k]_D= \frac {q_k} {Q} m_{3/2}^2 \cos^2 \theta
\frac{(n_{\Phi_2}-n_{\Phi_1})}2.
\ee

From this formula, it is apparent 
that the D-term contribution to the scalar
masses vanishes when the modular weights of $\Phi_1$ and $\Phi_2$ 
(or equivalently, their soft masses) are identical. If that is
the case, the vacuum is invariant under the interchange 
$\Phi_1 \leftrightarrow \Phi_2$ and hence
 $\langle \Phi_1 \rangle = \langle \Phi_2 \rangle$,
yielding $\langle D \rangle=0$ \footnote {This is a tree-level argument, 
and certainly, if $\Phi_1$ and $\Phi_2$ had different interactions,
radiative corrections could produce some differences
between $m^2_{\Phi_1}$ and $m^2_{\Phi_2}$ and the vacuum would not
be invariant under $\Phi_1 \leftrightarrow \Phi_2$ anymore. However, 
the effects of the running between the gravitational scale and
the $U(1)$ breaking scale are in general too small to have 
any significant effect. They might be relevant, though, for a
horizontal $U(1)$ symmetry, since the (flavour dependent) D-term 
contribution to the scalar masses represents 
a source of flavour violation.}.
The D-term also vanishes when
$\cos \theta =0$, i.e. the dilaton limit. This contrasts with
the anomalous $U(1)$ case, where the non-trivial transformation
of the dilaton with the anomalous symmetry translated into a contribution
to the scalar masses that depended on the dilaton auxiliary field,
generating a contribution even in the dilaton  
dominated SUSY breaking limit.
We conclude, then, that the 
dilaton limit can only be rescued 
when the light fields are charged under an anomalous gauge
group, broken at a high energy scale, and the charges 
are the appropriate ones. Nevertheless, when
 $\langle D \rangle $ is not vanishing, the D-term contribution
to the scalar masses is comparable to the soft masses, and
should be considered in the analyses.

As was mentioned in Section 2, if the only fields charged under
the SM group are the MSSM fields, there is only one anomaly-free
$U(1)$, apart from the weak hypercharge.
Under that symmetry, denoted by
$U(1)_H$, the Higgs doublets transform with 
opposite charges, while the rest of the particles
are neutral. In particular, the sleptons are neutral, so it
is not possible to lift the UFB-3 direction by the D-term
contribution to their scalar masses, as we did in the previous
section.
Furthermore, the fact that the Higgs doublets have opposite charges implies
that either $m^2_{H_1}$ or $m^2_{H_2}$ could be negative at the high
energy scale, with the subsequent potential problems
for the electroweak symmetry breaking. Nevertheless, 
this scenario deserves a careful analysis, since
a departure of universality in the Higgs masses could help to
rescue points from the UFB-3 bound. To be precise, if $H_2$
is heavier than the rest of particles, at the appropriate scale
to compute the effective potential, $m^2_{H_2}$ is less negative
than in the universal case. Therefore, the UFB-3 bound is alleviated.

We study in detail the case with $q_{H_2}=-q_{H_1}=  Q$, and 
$n_{\Phi_2}-n_{\Phi_1}=1$. We also fix all the modular weights
of the MSSM fields to $-1$, so that the only effects of 
non-universality come from the D-terms. That assignment of charges
produces a sparticle spectrum where, out of the dilaton limit,
$H_2$ has the largest soft mass and $H_1$ the smallest.
The departure of 
universality opens some small allowed windows, as is apparent 
from Figure 6. It should be noted that in the plots we have
varied the goldstino angle between 0 and $\pi$, and thus 
positivity is not guaranteed in the whole parameter space 
(positivity requires $\cos^2 \theta \ge 2/3$).
As a matter of fact, in
the allowed regions, the $H_1$ soft mass squared is negative. 
The reason is that the Higgs masses satisfy 
at high energies the relation 
$m^2_{H_1}+m^2_{H_2}=2 m^2_{rest}$, where $m^2_{rest}$ is the
common soft mass of the rest of the particles.
Then, had we required positivity of all the scalar masses squared,
the condition $m^2_{H_1} \geq 0$  would have translated
into $m^2_{H_2} \leq 2 m^2_{rest}$, and this 
non-degeneracy is not strong enough to lift the 
UFB-3 direction. As a result, the regions where all the scalar
masses squared are positive are excluded. Allowing
 $m^2_{H_1}$ to be negative,  $m^2_{H_2}$ can be very large, 
thus relaxing  the UFB-3 bound. 
On the other hand, and as was mentioned before,
if the Higgs masses squared
are  already negative at high energies, it is more difficult to 
fulfill the bound eq.(\ref{UFB-1}). However, there are 
some small windows where $m^2_{H_1}<0$, but 
the electroweak symmetry is broken 
in the correct way, and the non-degeneracy is strong enough
to lift the UFB-3 direction.

\begin{figure}
\begin{center}
\epsfig{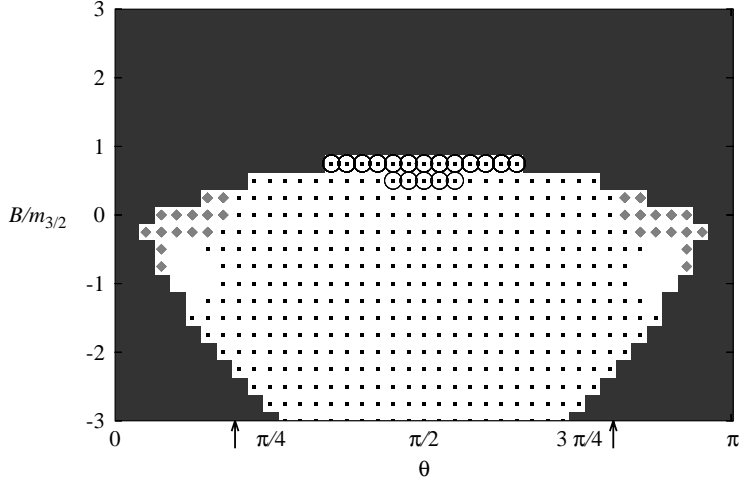}
\caption{
Excluded regions for a scenario where the Standard Model 
gauge group has been extended
with the non-anomalous $U(1)_{H}$ symmetry. The arrows
indicate the limits of the regions where 
positivity is not satisfied. 
The gravitino mass has been fixed to 100 GeV and
the meaning of the symbols is the same as in Figure 1.}
\end{center}
\end{figure}

\section{Application to realistic string models}

In the previous sections we have studied 
``string inspired'' scenarios, with the
SM gauge group extended to include one extra Abelian symmetry, 
anomalous or non-anomalous, and the matter content of the
MSSM plus some additional fields, to break the extra group. 
However, in a realistic model,
 there appear several extra $U(1)$s, one of 
them usually anomalous, and many chiral fields. Those extra
symmetries are broken  when certain
 fields acquire a vacuum expectation value, yielding
at low energies the SM group, the fields of the MSSM
and some other fields that are not observable, either because
they are very heavy, or because they are not charged under
the SM gauge group, i.e. they belong to the hidden sector. 
In this section, we will analyze some of these realistic models
to illustrate the impact of the D-term contribution to the 
scalar masses in a model with several $U(1)$s and several fields
acquiring vacuum expectation values, as happens generically
in explicit string constructions.

The gauge group in realistic models is of the form
 $SU(3)_C \times SU(2)_L \times U(1)^{N+1} \times ...$. 
Usually, there is a combination of those $U(1)$s which
is anomalous, $U(1)_A$, with vector superfield $V_A$, and
whose anomalies are cancelled by the Green-Schwarz mechanism. On
the other hand, the remaining $U(1)$s are non-anomalous, and the
corresponding vector superfields are $V_a$, $a=1,...N$. Concerning
the matter content, there appear many chiral superfields, $\phi_k$, 
that have a charge $q^A_k$ under $U(1)_A$
and $q^a_k$ under $U(1)_a$. Then, for orbifold compactifications, 
the tree level K\"ahler potential reads:
\be
\label{kahler-gen}
K=-\log (S+S^*-\delta_{GS} V_A)-3 \log(T+T^*)+
(T+T^*)^{n_{\phi_k}} 
e^{2 (q^A_k V_A+\sum_a q^a_k V_a)}  \phi_k^* \phi_k,
\ee
where we have omitted the non-Abelian part. 

To obtain the correct phenomenology, only one of the
$N$ non anomalous $U(1)$s should remain unbroken at
low energies, and  should
be identified with the weak hypercharge. Also, both the anomalous 
and the other $N-1$ non-anomalous combinations must be broken
along F- and D- flat directions, 
to preserve supersymmetry, when certain
scalars acquire a vacuum expectation value. We will denote 
these fields by $\Phi_k$. After the breaking of the 
 extra $U(1)$ symmetries, the scalar masses 
receive a contribution from the D-terms given by,
\be
\label{m-D-gen}
[m^2_k]_D=-q^A_k \langle D_A \rangle 
- \sum_a q^a_k \langle D_a \rangle,
\ee
where
\be
\label{vev-D-anom-gen}
\langle D_A \rangle= - m_{3/2}^2
 \frac {\sum_k Q^A_k  \langle \Phi_k \rangle ^2
(1-6 \sin^2 \theta - n_{\Phi_k} \cos^2 \theta)} 
{\sum_k (Q^A_k)^2  \langle \Phi_k \rangle ^2},
\ee
and
\be
\label{vev-D-noanom-gen}
\langle D_a \rangle= m_{3/2}^2 
 \frac {\sum_k Q^a_k  \langle \Phi_k \rangle ^2  n_{\Phi_k} \cos^2 \theta} 
{\sum_k (Q^a_k)^2  \langle \Phi_k \rangle ^2}.
\ee

The scalar masses depend crucially on the charges of the fields
involved, and on the vacuum expectation value of the fields that
break the $U(1)$s. In the next subsections, we analyze in
some detail two explicit string constructions that give definite
predictions for these quantities.

\subsection{Model I}

First, we analyze the model presented 
in \cite{Casas:1989us}, obtained from
the compactification of the heterotic string on the $Z_3$ orbifold
with Wilson lines \cite{Dixon:1985jw,Ibanez:1987tp}. This model
has several drawbacks that make it phenomenologically
unacceptable, but is a good example of what one expects
in a realistic string model: several $U(1)$s and several fields that
acquire a vacuum expectation value to break the extra groups.

At the gravitational scale, the model corresponds to the one presented in
\cite{Ibanez:1987sn}, with the gauge group $[SU(3) \times SU(2) \times
U(1)^5] \times [SO(10) \times U(1)^3]_{hidden}$. The spectrum of massless
particles consists on three generations of 76 fields, labeled $f_k$,
charged under the gauge group of the theory. 
Remarkably enough, it is possible to find  D- and F- 
flat directions where the gauge group is just 
$[SU(3) \times SU(2) \times U(1)_Y] \times [SO(10)]_{hidden}$, i.e. the 
SM group times a hidden sector group. In particular, this happens
for the following direction:
\bea
\label{vevs_1}
|\langle f_{12} \rangle |^2&=&2|\langle f_{58}\rangle|^2=
2|\langle \alpha \rangle|^2   \;\; ,\nonumber\\
|\langle f_{15}\rangle|^2&=&|\langle f_{18}\rangle|^2=
3|\langle \alpha \rangle|^2 \;\; ,\nonumber\\
|\langle f_{21}\rangle|^2&=&2|\langle f_{32}\rangle|^2=
6|\langle \alpha \rangle|^2 \;\; ,\nonumber\\
|\langle f_{43}\rangle|^2&=&
4|\langle \alpha \rangle|^2 \;\; ,\nonumber\\
|\langle f_{54}\rangle|^2&=&|\langle f_{65}\rangle|^2=
|\langle \alpha \rangle|^2 \;\;,
\eea
where $|\langle \alpha \rangle|^2=\sqrt{3}/24 \; \xi^2$.
Furthermore, after the breaking of the extra symmetries, many
of the 76 fields become heavy, while others remain massless.
The spectrum of massless particles is very similar to the MSSM:  
$3 \{(3,2)+2(\bar{3},1)+(1,2)+(1) \}+
3\{4(1,2)+4 (1)+(16)'+11 (1)'\}$, where the prime
denotes a representation that is invisible at low
energies. The only differences are that there are
three generations of Higgses, $3 \times 2$ extra doublets
and $3 \times 4 $ extra singlets. These fields have not
been observed, and so this model is not completely realistic.
However, it has some non-trivial predictions, namely 
${\cal N}=1$ SUSY, the SM gauge group, three generations, and a matter
content that resembles the MSSM. The similarity of this
model with the MSSM is remarkable and suggests that the
true vacuum might not be very different from the one
predicted. In any case, we will use this ``semi-realistic'' 
string model to illustrate the impact of the D-term contribution
to the scalar masses on the CCB and UFB bounds.

From the charges of the different fields and eq.(\ref{vevs_1}) it is
straightforward to compute the scalar masses of the MSSM fields
at high energies. In this model, all the fields that acquire a 
vacuum expectation value belong to the twisted sector, and thus
have the same modular weight, $-2$. Hence,
the contribution to the scalar masses from the breaking of
the non-anomalous $U(1)$s vanishes and the only contribution comes
from $U(1)_A$. 
On the other hand, regarding
the MSSM fields, $Q_L$, $u_R$ and $H_2$ belong to the untwisted
sector (modular weight $-1$), and the rest, to the twisted sector
(modular weight $-2$). 
The charges of the relevant fields are:
\be
\label{CKM1}
\begin{tabular}{c|ccccccccc}
    & $f_{12}$ & $f_{15}$ & $f_{18}$ & $f_{21}$ &
     $f_{32}$ & $f_{43}$ & $f_{54}$ & $f_{58}$ & $f_{65}$ \\
\hline
   $Q^A$& $\frac{-8}{\sqrt{3}}$  &  $\frac{-8}{\sqrt{3}}$  & 
$\frac{-8}{\sqrt{3}}$  &
 $\frac{-8}{\sqrt{3}}$  &  $\frac{4}{\sqrt{3}}$  & 
$\frac{-8}{\sqrt{3}}$  &
$\frac{-8}{\sqrt{3}}$  & $\frac{4}{\sqrt{3}}$  & 
$\frac{-8}{\sqrt{3}}$ 
\\
\end{tabular}
\ee
\be
\label{CKM2}
\begin{tabular}{c|ccccccc}
 & $Q_{L}$ & $d_R$ & $u_R$ & $L_{L}$ & $e_R$ 
& $H_1$ & $H_2$ \\
\hline
$q^A$ &0 & $\frac{4}{\sqrt{3}}$ & 0 & $\frac{4}{\sqrt{3}}$ &
 $\frac{-8}{\sqrt{3}}$ &
$\frac{4}{\sqrt{3}}$ & 0
\\
\end{tabular}
\ee
Therefore, the scalar masses are:
\bea
\label{masses_1}
&m^2_{Q_L}= m^2_{u_R}= m^2_{H_2}= m^2_{3/2}(1-\cos^2 \theta) \;\; ,&\nonumber\\ 
&m^2_{d_R}=m^2_{L_L}=m^2_{H_1} =
 m^2_{3/2} \left(\frac{3}{7}-\frac{19}{7} \cos 2 \theta\right)\;\; ,&\nonumber\\ 
&m^2_{e_R} =   m^2_{3/2} \left(-\frac{6}{7}+\frac{17}{7} \cos 2 \theta \right)\;\; .&
\eea
From these formulas it can be checked that at high energies there
is always a charged field whose mass squared is negative, thus
breaking charge. The problematic field is $e_R$: it has a charge
under the anomalous $U(1)$
of opposite sign to the charges of the rest of the MSSM particles. 
Incidentally, this field
is also problematic in the model presented in  \cite{Casas:1989us},
because there are three singlets
 under $SU(3) \times SU(2)$ with hypercharge 1, and any of them could
 play the role of the $e_R$ (of course, one should find the
way to make the other two heavy to make the model realistic). It
is worth noticing that one of the possible $ e_R$s, 
the combination of $\{f_8,f_{25},f_{39},f_{42},f_{49},f_{59},
f_{66},f_{75},f_{76}\}$, has charge $4/\sqrt{3}$ under $U(1)_A$,
identical to the charges of $d_R,L_L,H_1$. If we identify
 $e_R$ with this combination of fields, there will be some
ranges of $\theta$ where all the scalar masses will be positive.
If we do that, the scalar masses of the MSSM fields  read:
\bea
\label{masses_1_p}
&m^2_{Q_L}= m^2_{u_R}= m^2_{H_2}= m^2_{3/2}(1-\cos^2 \theta)
 \;\; ,&\nonumber\\ 
&m^2_{d_R}=m^2_{L_L}=m^2_{H_1} =m^2_{e_R}=
 m^2_{3/2} \left(\frac{3}{7}-\frac{19}{7} \cos 2 \theta\right)\;\;, &
\eea
which are simultaneously positive for $\cos 2 \theta < 3/19$.
The positivity of these scalar masses at high energies is a
necessary condition for the phenomenological viability of
the model. However, one must also impose, among other things,
a correct top mass, SUSY masses above the experimental bounds, 
and the absence of CCB minima and UFB directions. 
To perform the analysis, we  also need to compute
the trilinear soft terms. This has the
same limitation as in the ILR scenario (see Sect. 2.3), i.e. some
of the fields belong to the twisted sector and the trilinear
couplings could become T-dependent. Nevertheless, for the UFB-3 bound,
the most important one for our analysis, only the top trilinear
term is relevant, and the corresponding fields, 
$Q_L,u_R$ and $H_2$, are all untwisted in this model. 
Consequently, the limitation mentioned above does not apply and 
the trilinear term is simply 
$A_{t}=\sqrt{3} m_{3/2} \sin \theta$.

It is interesting to note
that in the dilaton limit the slepton masses are approximately 
three times larger than the $H_2$ mass, which helps to
prevent the appearance of UFB directions. Therefore, one
expects to find some regions in the parameter space where 
all our phenomenological requirements are satisfied (incidentally,
if we had ignored the D-term contribution to the scalar masses,
the whole parameter space would be excluded). This
is confirmed by the numerical analysis, shown in Figure 7. If we
depart from the dilaton limit, it
is still possible to find allowed regions in the parameter space. For
example, for a gravitino mass of 500 GeV, there are allowed regions
for $1.3 \lsim \theta \lsim 1.8$.

\begin{figure}
\begin{center}
\epsfig{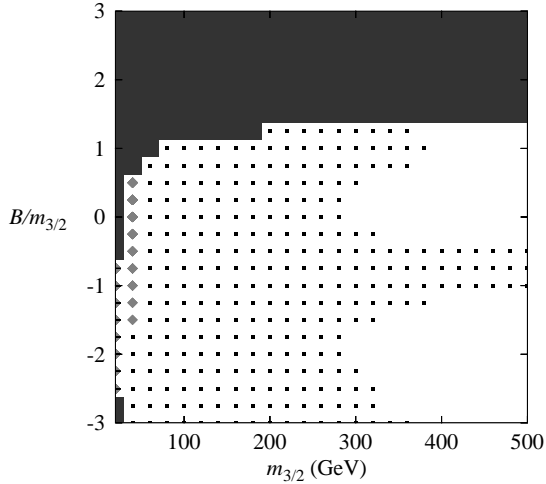}
\caption{Allowed regions for the Model I, presented in 
\cite{Casas:1989us}, assuming SUSY breaking dominated by the dilaton.
The meaning of the symbols is the same as in Figure 1.}
\end{center}
\end{figure}

\subsection{Model II}

We analyze now the model presented
 in \cite{Faraggi:1990ka,Cleaver:1999sa,Cleaver:2001cj},
 constructed in
the four-dimensional free fermionic 
string formulation \cite{Antoniadis:1987rn}. This model
has several remarkable properties. First, all 
the exotic fractionally charged states and all the color triplets
receive a mass of the order of the string scale. Furthermore,
even though initially there are four fields with the same quantum
numbers as the up and down Higgs,
it is possible to find a flat direction where only one Higgs doublet
remains light. At the end of the day, we obtain at low energies
a model with just the MSSM spectrum and the gauge group $SU(3)_C 
\times SU(2)_L \times U(1)_Y \times G_{hidden}$. It is equally
remarkable that the
weak hypercharge has the standard $SO(10)$ normalization.

Let us now briefly describe the model. At high energies, the 
gauge group is $SU(3)_C \times U(1)_C \times SU(2)_L \times U(1)_L
\times U(1)^6$ in the observable sector, and $SO(4) \times
SU(3) \times U(1)^4$ in the hidden sector. There are six anomalous
$U(1)$s, although it is possible to combine them to yield
five non anomalous and one anomalous $U(1)$. The corresponding
Fayet-Iliopoulos term must be compensated by the VEVs of some fields
to preserve supersymmetry. If those fields are also charged under
other $U(1)$s, these other groups will also be broken. The VEVs
of all the fields must satisfy the condition that all the 
D-terms and the F-terms are cancelled simultaneously, to preserve
supersymmetry, and that all the $U(1)$s in the observable sector
are broken except one combination, which should be identified with
the weak hypercharge. Obviously, it is highly non-trivial to fulfill
all these conditions. However, this model presents several flat 
directions where this actually happens. For example, 
the direction defined by
\bea
\label{vevs_2} 
&(\langle{\phi_4}\rangle \langle{{\bar\phi}_4'}\rangle +
\langle{{\bar\phi}_4}\rangle \langle{\phi_4'}\rangle )=0 \;\; , &\nonumber \\
&{1\over 3}| \langle{\phi_{12}}\rangle | ^2=
            | \langle{\phi_{23}}\rangle | ^2=
            | \langle{{\bar\phi}_{56}}\rangle | ^2
= |  \langle{\alpha}\rangle  | ^2  \;\; ,&\nonumber \\
&  {1\over 2}| \langle{H_{15}}\rangle | ^2=
  {1\over 3}| \langle{H_{30}}\rangle | ^2=
            | \langle{H_{31}}\rangle | ^2=
  {1\over 2}| \langle{H_{38}}\rangle | ^2
= |  \langle{\alpha}\rangle  | ^2  \;\; ,&\nonumber \\
& (| \langle{\phi_4}\rangle | ^2+
| \langle{\phi_4^\prime}\rangle | ^2)-
(| \langle{{\bar\phi}_4}\rangle | ^2+
| \langle{{\bar\phi}_4^\prime}\rangle | ^2)=
|  \langle{\alpha}\rangle  | ^2 \;\; ,&
\eea
where $|\langle \alpha \rangle|   \approx 7\times 10^{16}$ GeV,
only leaves the combination $\frac{1}{3}U(1)_C +\frac{1}{2}U(1)_L 
\equiv U(1)_Y$ unbroken \cite{Cleaver:1999sa}.

Concerning the matter content, at high energies the massless spectrum 
includes fields both from the Neveu-Schwarz (NS)
 sector and the  ${\mathbf b_1}$, ${\mathbf b_2}$,
${\mathbf b_3}$ sectors.
The NS sector gives rise to the spin-2 states, the gauge bosons in the observable and
hidden sectors, and scalar representations belonging to the
observable sector (denoted by $h$ and $\phi$). These fields are
untwisted and hence have modular weight $-1$. 
On the other hand, the ${\mathbf b_1}$, ${\mathbf b_2}$,
${\mathbf b_3}$ sectors give
 rise to the (twisted) chiral fields 
of the observable sector (denoted by $V$, $H$ and the three
families of quarks and leptons). In the 
free fermionic models there is an underlying $Z_2 \times Z_2$
orbifold, so, the modular weights of the 
twisted fields are also $-1$  \cite{Ibanez:1992hc}.

The scalar masses for this model can be readily computed
from eqs.(\ref{m-D-gen},\ref{vev-D-anom-gen},\ref{vev-D-noanom-gen}).
All the fields that acquire a VEV have modular weight --1,
then, only the breaking of the anomalous $U(1)$ has an
impact on the scalar masses --- the non-anomalous contribution 
vanishes. The charges of those fields under the
anomalous $U(1)$ symmetry are:
\be
\label{FNY1}
\begin{tabular}{c|ccccccccccc}
   & $\phi_{12}$ & $\phi_{23}$ & $\bar \phi_{56}$ &
$H_{15}$ & $ H_{30}$ & $H_{31}$ & $H_{38}$ & $\phi_4$ &
$\phi'_4$ & $\bar \phi_4$ & $\bar \phi'_4$ \\
\hline
$Q^A$  &--9 & --2 & --2 & --2 & --6 & 3 & --4 &
0 &0 & 0 &0\\
\end{tabular}
\ee

Notice that the conditions eqs.(\ref{vevs_2})
do not fix the VEVs of the 
$\{ \phi_4,\phi'_4,\bar \phi_4,\bar \phi'_4 \}$ fields to 
any particular value. However, this is not a problem for
the calculation of the scalar masses, because those fields
are neutral under $U(1)_A$. Then, a straightforward
calculation yields
\bea
\label{massesFNY}
m^2_k=m^2_{3/2}\left(1+\frac{145}{204}q_k \right) \sin^2 \theta,
\eea
where $k$ runs over all the MSSM fields. The scalar soft masses depend
both on the goldstino angle and the charges under $U(1)_A$. These are:
\be
\label{FNY2}
\begin{tabular}{c|ccccccc}
$\alpha$ & $Q_{L_{\alpha}}$ & $d_{R_{\alpha}}$ & 
$u_{R_{\alpha}}$ & $L_{L_{\alpha}}$ & $e_{R_{\alpha}}$ 
& $h_{\alpha}$ & $\bar h_{\alpha}$\\
\hline
 1 & 2 & 2 & 2 & 2 & 2 & 4   & --4 \\
 2 & 3 & 2 & 3 & 2 & 3 & --5 &   5 \\
 3 & 2 & 1 & 2 & 1 & 2 & --3 &   3 \\
\end{tabular}
\ee
where $\alpha$ indicates the sector $b_{\alpha}$ which gives rise
to the field (the index  $\alpha$ is not related a priori with 
the generation of the field).
There are two more  MSSM-like fields, 
$H_{34}$ and $H_{41}$, that have the same charges 
under the SM gauge group as 
the up and down Higgs doublets, and charges 2 and 0 under $U(1)_A$,
respectively.
In principle, the presence of four light Higgs generations
would be a problem for the phenomenological 
viability of the model (the
gauge couplings would not unify, for example). 
However the $4 \times 4$ mass matrix yields only
one massless pair, $(-h_1+\sqrt{3} h_3)/2 \equiv H_1$ and
$\bar h_1 \equiv H_2$, and three with a mass of the order of the 
string scale. 

The renormalizable part of the 
MSSM superpotential reads:
\be
\label{FNY-W}
W_{MSSM}=g_s \sqrt{2} H_2 Q_1 u_{R_1} +g_s \sqrt{2} \cos \sqrt{3}
H_1 [Q_3 d_{R_3}+L_3 e_{R_3}],
\ee
where we have omitted a term involving right handed neutrinos,
which is irrelevant for our discussion.
The quark and lepton fields in eq. (\ref{FNY-W}) 
should be identified with the third generation particles
 (for example, $u_{R_1} \equiv t_R$), whereas the two
first generations only appear in the superpotential through
non-renormalizable terms of order four or six.

It is apparent from eq.(\ref{massesFNY}) that if some field
has a $U(1)_A$ charge $\lsim -3/2$, the corresponding 
scalar mass squared will be negative at high energies. It is interesting that
squarks and sleptons have positive charge under $U(1)_A$ and therefore
their masses squared are positive. 
However, this is not true for the Higgs doublet $H_2$,
whose mass squared is negative at high energies. 
This result is a consequence of the gauge invariance of
the superpotential under $U(1)_A$.  Gauge invariance requires that,
in every term of the superpotential, at least one
of the fields has a negative charge. If there is a
term involving only MSSM fields,  some of the MSSM fields
will have negative charge. As a matter of fact, 
this is what happens in this model,
where a Yukawa coupling involving $H_2$ and other MSSM fields 
appears at the renormalizable level. 
Then, some of the MSSM fields must have negative
charge and the corresponding scalar masses squared are negative
at high energies. Things are different, though, for the other
Higgs doublet, $H_1$. $H_1$ does not have a definite $U(1)_A$
charge, but is a mixed state of fields with different charges: 
$h_1$, with charge 4, and $h_3$, with charge $-3$.
The component that couples to the MSSM fields in the renormalizable
superpotential is $h_3$, whose  
mass squared is negative. However, we are interested in
$m^2_{H_1}$, that is a combination of the masses squared of 
$h_1$ and $h_3$. The former is positive and the latter is negative,
but $m^2_{H_1}$ turns out to be positive. This shows that it
is possible to have a renormalizable superpotential involving
only MSSM fields, and at the same time all the scalar masses 
squared positive at high energies, provided the Higgs doublets
$H_1$ and $H_2$ are mixed states with a certain component of
a field with positive $U(1)_A$ charge \footnote{
A second way of getting positive masses squared 
is in a model where the MSSM Lagrangian
comes from a non-renormalizable superpotential, as the models in 
Sections 2 and 3. Then,
gauge invariance allows a term in the superpotential in
which all the MSSM fields have a positive charge, and the negative
charge is carried by a field or fields that will 
eventually acquire a VEV.}.

As was mentioned in Section 2, negative Higgs masses squared
represent a potential problem for a correct electroweak symmetry
breaking, but they do not prevent it: the contribution to the
Higgs masses from the $\mu$ term can lift the UFB direction
in the Higgs potential that would  otherwise  appear. Therefore,
we expect that in some regions of the parameter space
the electroweak symmetry breaking occurs 
in the usual way. In Figure 8 
we show the forbidden regions in the parameter 
space of this model, for a
gravitino mass of 100 GeV. From the figure it can be seen that
it is possible to break the electroweak symmetry generating
a top mass of 174 GeV, but the global minimum always breaks charge.
The reason is that the UFB-3 direction is particularly dangerous
in this model. First, because $m^2_{H_2}$ is large and negative,
and second, because $m^2_L$ is small and cannot compensate the
negative contribution from  $m^2_{H_2}$
in the first term of eqs.(\ref{SU8},\ref{SU9}). 

\begin{figure}
\begin{center}
\epsfig{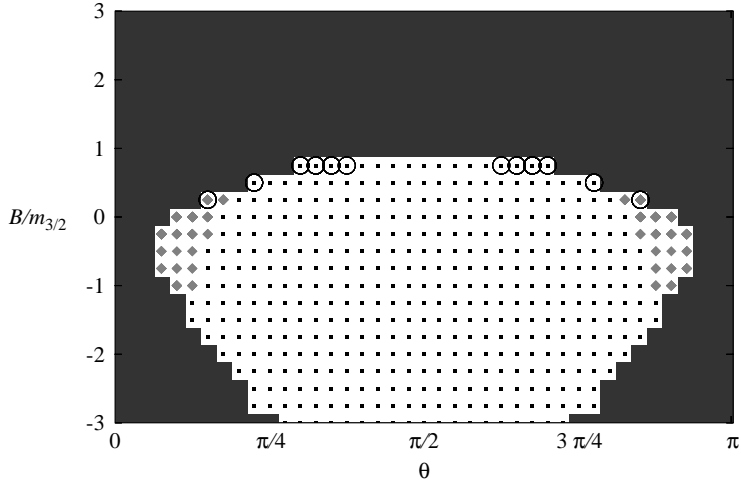}
\caption{Forbidden regions for the Model II, presented in 
\cite{Faraggi:1990ka,Cleaver:1999sa,Cleaver:2001cj}, for a gravitino
mass of 100 GeV.
The meaning of the symbols is the same as in Figure 1.}
\end{center}
\end{figure}

\section{Conclusions}

We have performed an analysis of the parameter space of different
weakly coupled string scenarios, imposing 
the requirements of an acceptable electroweak symmetry breaking,
a top mass within the experimental range, 
supersymmetric masses above the experimental 
bounds, and the absence of dangerous 
charge and color breaking (CCB) minima or 
unbounded from below (UFB) directions.
We have taken into account the fact that in superstring models
there are usually extra $U(1)$s broken at a high energy scale,
that contribute to the scalar masses through the vacuum expectation
value of the $D$ field associated to the $U(1)$ vector superfield.
Then, the spectrum of supersymmetric particles can be very different
to the one obtained in a superstring scenario with just the Standard Model
gauge group, where the only contribution to the scalar soft masses
comes from the vacuum expectation value of the dilaton and/or moduli 
auxiliary fields.

We have shown that the presence of family-independent 
 extra $U(1)$s can help
to lift the unbounded from below directions that appear in
the scalar potential when the  gauge group is just the Standard Model 
gauge group. To be precise, this happens
 when the charges of the fields under the $U(1)$s
are such that the slepton masses are larger than the rest. In 
particular, we have studied a string motivated scenario where the gauge
group of the Standard Model is extended with an 
extra anomalous (non-anomalous) $U(1)$ and the matter
content of the MSSM is extended with one (two) chiral superfields to break
the $U(1)$. 
In the anomalous case, we have found 
that the non-universality in the scalar masses,
generated by the breaking of the anomalous $U(1)$, can be large enough
to open some allowed windows in the parameter space. We have also
studied some realistic string scenarios, 
with several extra $U(1)$s, anomalous
and non-anomalous, and several fields that 
acquire a vacuum expectation value.
Again, we find that the D-term contribution 
to the scalar masses is crucial
to lift the UFB directions that otherwise appear.

We have analyzed in detail the case where the 
dilaton is the field that dominates the supersymmetry breaking.
Since the dilaton couples universally to all the particles, 
the contribution to the scalar masses from the dilaton $F$-term
is universal, thus alleviating the supersymmetric flavour and CP problems.
If the gauge group at high energies 
is just the Standard Model, this scenario is essentially
ruled out because the global minimum of the 
effective potential breaks charge. 
However, in a realistic string theory there are usually extra $U(1)$s
broken at high energies that modify the spectrum of scalar masses. 
We have seen that the non-anomalous $U(1)$s do not contribute to the
scalar masses in the dilaton limit, but the anomalous $U(1)$ does.
Depending on
the charges of the particles under the anomalous $U(1)$, it is possible
to find regions of the parameter space  where our
physical vacuum is the global minimum of the effective potential,
namely, when the charges are such that the slepton masses are large.
At the same time, if the extra gauge group is family blind,
this scenario keeps all the desirable properties mentioned above, 
regarding the flavour and CP problems in supersymmetric theories.
It is indeed remarkable that there are models
with allowed windows in this interesting limit, 
which has important implications
for flavour and CP violation physics.

\subsection*{Acknowledgments}

The author would like to thank  Alon Faraggi, 
Carlos Mu\~noz, Liliana Velasco-Sevilla, and
especially Alberto Casas and Graham Ross 
for invaluable suggestions and a
careful reading of the manuscript. The author also thanks
Biagio Lucini for help with the computers and PPARC for
financial support.

\end{document}